\documentclass[pre,twocolumn]{revtex4-2}
\usepackage{graphicx} 
\usepackage{amsmath}

\usepackage{color}
\usepackage[colorlinks=true,citecolor=blue,urlcolor=blue,linkcolor=blue]{hyperref}

\usepackage[mathlines]{lineno}
\let\oldequation\equation\let\oldendequation\endequation
\renewenvironment{equation}{\linenomathNonumbers\oldequation}{\oldendequation\endlinenomath}
\let\oldalign\align\let\oldendalign\endalign
\renewenvironment{align}{\linenomathNonumbers\oldalign}{\oldendalign\endlinenomath}

\begin{document}
   \title{Circling crystals in chiral active matter with self-alignment}

    \author{Marco Musacchio}
    \email{musacchio@thphy.uni-duesseldorf.de}
	\affiliation{
		Institut f{\"u}r Theoretische Physik II: Weiche Materie,
		Heinrich-Heine-Universit{\"a}t D{\"u}sseldorf, Universit{\"a}tsstra{\ss}e 1,
		D-40225 D{\"u}sseldorf, 
		Germany}
    
	\author{Alexander P.\ Antonov}
	\affiliation{
		Institut f{\"u}r Theoretische Physik II: Weiche Materie,
		Heinrich-Heine-Universit{\"a}t D{\"u}sseldorf, Universit{\"a}tsstra{\ss}e 1,
		D-40225 D{\"u}sseldorf, 
		Germany}
	
	\author{Hartmut L{\"o}wen}
	\affiliation{
		Institut f{\"u}r Theoretische Physik II: Weiche Materie,
		Heinrich-Heine-Universit{\"a}t D{\"u}sseldorf, Universit{\"a}tsstra{\ss}e 1,
		D-40225 D{\"u}sseldorf, 
		Germany}

    \author{Lorenzo Caprini}
    \email{lorenzo.caprini@uniroma1.it}
	\affiliation{
		Physics Department, University of Rome La Sapienza, P.le Aldo Moro 5, IT-00185 Rome, Italy}
	
	\date{\today}
        
	\begin{abstract} 
       \noindent 
       We study a crystal composed of active units governed by self-alignment and chirality. The first mechanism acts as an effective torque that aligns the particle orientation with its velocity, while the second drives individual particles along circular orbits. We find that even a weak degree of chirality, when coupled with self-alignment, induces collective motion of the entire crystal along circular trajectories in space. We refer to this phase as a \textit{circling crystal}.
       When chirality outweigh self-alignment, the circular global motion is suppressed in favor of vortex-like regions of coordinated motion. This state is characterized by oscillating spatial velocity correlations, a power law decay of the energy spectrum, and oscillatory temporal correlations. Our findings can be tested experimentally in systems ranging from epithelial tissues to swarming robots, governed by chirality and self-alignment.
       \end{abstract}
	
	\maketitle
    
\makeatletter
\def\widetext{%
  \par
  \vskip\intextsep
  \onecolumngrid
  \vspace*{-\baselineskip}
}
\def\endwidetext{%
  \par
  \vskip\intextsep
  \twocolumngrid
  \vspace*{-\baselineskip}
}
\makeatother

\section{Introduction}
\label{sec:introduction}
In the last years, active matter~\cite{marchetti2013hydrodynamics, Elgeti2015review, bechinger2016active} has emerged as an exciting new field for applying and testing concepts of non-equilibrium physics. Active systems are typically composed of self-propelled units that convert energy from internal or external sources into directed motion. These units often tend to follow circular or helical orbits, systematically rotating clockwise or counterclockwise. This kind of systems are often termed \textit{chiral}~\cite{Löwen2016_ChiralityMicroswimmer, LiebchenLevis:2022} because their trajectory cannot be superimposed by their mirror images. Examples of chiral active matter are widespread in biology: they range from living microswimmers, such as bacteria~\cite{DiLuzioEtAl:2005RightHandSide, wioland2013confinement, petroff2015fast, perez2019bacteria} moving on a substrate to malaria parasites~\cite{patra2022collective}, chiral microtubules driven by molecular motors~\cite{afroze2021monopolar}, sperm cells~\cite{Riedel2005SelfOrganized} and small cell clusters confined in circular islands~\cite{wang2022cellchirality}.
Chirality has also been implemented in the design of artificial systems by breaking the rotational symmetry in the body shape or in the self-propulsion mechanisms, as in the case of L-shaped colloids~\cite{KummellEtAl:2013CircularMotion,shelke2019transition}, isotropic droplets swimming in circles due to chiral ordering and geometrical frustration~\cite{sevc2012geometrical, carenza2019rotation}, or chiral active granular particles~\cite{scholz2018rotating, chor2023manybody, Siebers2023Exploiting, Chan2024, CapriniEtAl:2025SelfWrapping, boriskovsky2025probing, KantEtAl:2025EdgeStates, carrillo2025depinning, noirhomme2025brainbots, Novkoski2025} often termed spinners~\cite{Workamp2018, scholz2021surfactants, LopezCastano2022ChiralFlow, LopezCastano2022ChiralityTransitions}.

The impact of chirality on single-particle properties, such as odd diffusion~\cite{hargus2021odd, Kalz2022Collisions, Kalz2024Oscillatory, Muzzeddu2025OddTracer} and edge currents in external potentials~\cite{Caprini2023camextpotential, WangBerx:2025BraidedMixing}, has been extensively explored, while recent efforts have focused on chirality-induced collective phenomena~\cite{Liao2021emergent, Kreienkamp2022_ClusteringFlockingNonReciprocal, Hiraiwa2022_CollisionInducedTorque, Ceron2023_Swarmalators, Lei2023, Mecke2024}. Both theoretical analyses and numerical simulations have shown that adding chirality suppresses motility-induced phase separation~\cite{liao2018clustering, bickmann2022analytical, Langford2025}, leading to the formation of hyperuniform fluids~\cite{lei2019nonequilibrium,zhang2022hyperuniform, kuroda2023microscopic} and the emergence of intriguing structures such as rotating vortices in the velocity field~\cite{zhang2020reconfigurable} or self-reverting vorticity~\cite{caprini2024self}. Introducing alignment interactions in chiral systems~\cite{kruk2020traveling, zhang2022collective, Negi2023, Wang2024, Jian2025ChiralPhase} leads to an enhanced flocking behavior and pattern formation~\cite{liebchen2017collective} or micro-flock pattern~\cite{levis2018microflock} due to the fact that the interplay between self-rotation and alignment fosters synchronization among particles and stabilizes large-scale coherent chiral motion.
A recent line of research focuses on chiral active crystals, where each fundamental unit of the solid consists of chiral active particles. In this case, chirality reduces the correlation length of the spatial velocity correlations~\cite{Shee2024Emergent,marconi2025spontaneous} typical of active matter and induces cross–spatial correlations among different velocity components, resulting in a net angular momentum~\cite{marconi2025spontaneous}. 
Moreover, odd-elastic or chiral crystals exhibit spontaneous grain rotation, autonomous propulsion, and dynamic fragmentation~\cite{huang2025anomalous}.
In addition, the chirality-induced suppression of displacement fluctuations associated with hyperuniformity generates truly long-translational order even in two dimensions~\cite{kuroda2025long}.

\begin{figure*}[t] 
    \centering
    \includegraphics[width=1.8\columnwidth]{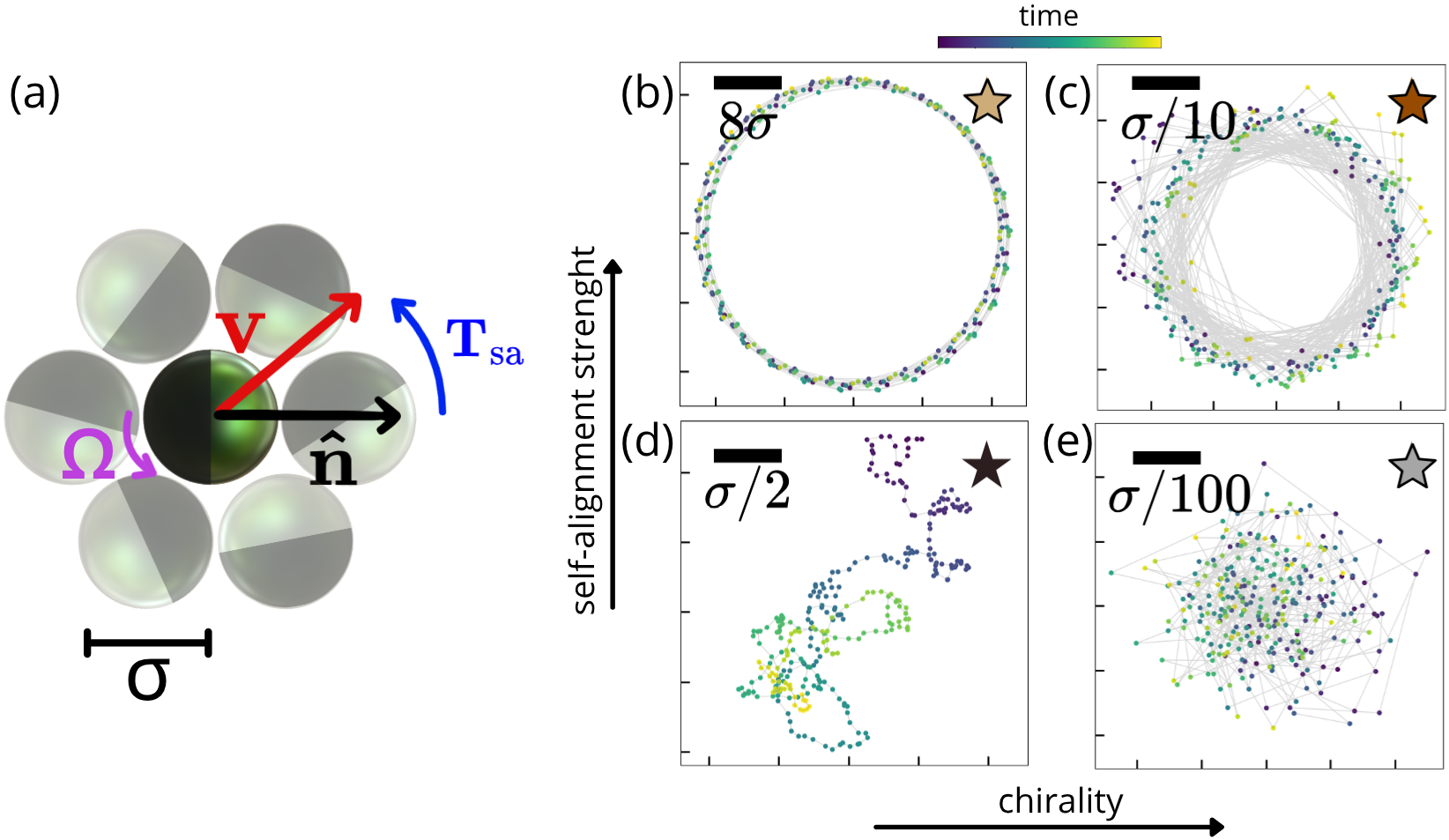}
    \caption{\textbf{Schematic and dynamics of chiral self-aligning active solids.} (a) Schematic illustration of an active solid composed of active particles characterized by self-alignment and chirality. Here, the particle orientation, indicated by the black cap on the particle, is represented by $\hat{\mathbf{n}}$, while $\mathbf{v}$ denotes its velocity. The rotation direction, i.e.\ the particle chirality, is indicated through a purple circular arrow, while the self-alignment torque $\mathbf{T}_{sa}$ by a blue circular arrow. (b-e) Trajectories of a target particle within the active solid as the strengths of self-alignment and chirality are varied. The color along the trajectory encodes time, with darker points representing earlier positions and lighter points indicating later positions. The points along each trajectory are connected by gray lines to highlight the particle’s path. The simulation results presented here were obtained for reduced self-alignment values of $B = 0.1$ and $B = 0.5$, and reduced chirality values of $\omega = 0.5$ and $\omega = 40$. The other dimensionless parameters of the simulations are: $\text{Pe} = 10$, $M = 10^{-4}$, $\sqrt{\epsilon/m}/(D_r \sigma)=10^2$, and $\Phi = N \pi \sigma^2 / 4L^2 = 1.1$.}
    \label{fig:figura1}
\end{figure*}

Several experimental systems, such as cells~\cite{szabo2006phase}, epithelial tissues~\cite{henkes2011active, barton2017active, giavazzi2018flocking}, and active granular particles~\cite{giomi2013swarming, dauchot2019dynamics,  tapia2021trapped, Carrillo-Mora2025}, are characterized by a coupling between translational and rotational motion commonly referred to as \textit{self-alignment}~\cite{baconnier2024self}. This mechanism acts as an effective torque that tends to align the particle’s orientation with its velocity~\cite{malinverno2017endocytic, Teixeira2021, baconnier2022selective, paoluzzi2024flocking, casiulis2025geometric, Musacchio2025, Kinoshita2025}. Self-alignment is one of the key mechanisms that can lead to flocking behavior~\cite{paoluzzi2024flocking, Musacchio2025} even if this is a single-particle mechanism that does not couple the orientation of different particles. 

The influence of self-alignment on collective phenomena has been recently investigated in several studies. 
Specifically, self-alignment suppresses motility-induced phase separation in favor of a homogeneous flocking phase~\cite{Musacchio2025}, and it is responsible for a re-entrant glass transition~\cite{paoluzzi2024flocking} due to emergent migratory patterns. In addition, recent studies focus on the case of self-aligning active crystals, where each unit of the solid is subject to a self-aligning activity. While, in the absence of self-alignment, active crystals excite collective modes at every wave vector~\cite{caprini2023entropons}, self-alignment activates a few elastic modes, as proved experimentally~\cite{baconnier2022selective} and theoretically~\cite{hernandez2024model}. In this case, the system displays flocking behavior which can be analytically predicted and interpreted as a second-order phase transition described by an effective velocity-dependent free energy with a Landau-Ginzburg profile~\cite{Musacchio2025pt}.

However, the interplay between chirality and self-alignment remains scarcely explored. It is still unclear if the competition between these two mechanisms influence the collective dynamics of active systems and generate different collective behavior. 

In this paper, we explore the competition between self-alignment and chirality on collective behaviors in dense active systems, with crystalline order. We discover that even a weak chirality introduces a global rotation, giving rise to a new collective phenomenon, which we term \textit{circling crystal}. In this phase, all the particles in the system are able to coordinate their motion and move collectively along circular trajectories. By contrast, strong chirality suppresses the global circling in favor of vortex structures in the velocity field. Here, the system is characterized by oscillating spatial velocity correlations with a finite swirl size and a power-law decaying energy spectrum, while the velocity auto-correlations show a time oscillating profile.

The structure of the paper is as follows: In Sec.~\ref{sec:model}, we present the model used to perform the simulations of a solid composed of active units characterized by self-alignment and chirality. The corresponding results are discussed in Sec.~\ref{sec:results}. Finally, in Sec.~\ref{sec:conclusions}, we provide a general discussion together with possible perspectives, applications, and directions for future studies and experimental validations. The details of the model employed in the numerical simulations are presented in Appendix~\ref{appendix:A}, while the derivations supporting our theoretical predictions are given in Appendix~\ref{app:theory}.

\section{Model}
\label{sec:model}


We consider a two-dimensional solid consisting of $N$ interacting chiral active Brownian particles subject to a self-alignment mechanism.
Particles are active and they move with a constant velocity $v_0$ along their intrinsic orientation $\hat{\mathbf{n}}$ (black arrow in Fig.~\ref{fig:figura1}(a)), indicated by the black cap on each particle, and are chiral since their orientation rotates at constant frequency, $\Omega$ (purple circular arrow in Fig.~\ref{fig:figura1}(a)), also termed chirality. This rotating mechanism pushes a free particle to move along a circular trajectory whose typical radius is determined by the interplay between the chirality strength and the particle's self-propulsion, $r_{\Omega}=v_0/\Omega$.
In this system, the crystal units are subject to a self-alignment mechanism, modeled as an effective torque, $\mathbf{T_{sa}}$ (blue circular arrow in Fig.~\ref{fig:figura1}(a)), that tends to align the particle orientation, $\mathbf{\hat{n}}$, with its velocity $\mathbf{v}$ (red arrow in Fig.~\ref{fig:figura1}(a)).
%
%
To evolve the particles dynamics, we consider an inertial translational dynamics and an overdamped rotational dynamics, given by
\begin{subequations}
\label{eq:dynamics}
    \begin{align}
    m\dot{\mathbf{v}}_i &= -\gamma \mathbf{v}_i + \gamma v_0\mathbf{\hat{n}}_i + \mathbf{F}_i + \gamma\sqrt{2 D_t}\,\boldsymbol{\xi}_i \,, \label{eq:eulero_pos} \\
     \gamma_r \dot{\theta}_i  &=  \gamma_r \Omega + \mathbf{T}^{sa}_i \cdot \hat{\mathbf{e}}_z + \gamma_r \sqrt{2 D_r} \eta_i,  \,
     \label{eq:eulero_ang}
\end{align}
\end{subequations}
where $\mathbf{x}_i$ is the position of the \textit{i} particle, $\mathbf{v}_i = \dot{\mathbf{x}}_i$ is its velocity, and $\theta_i$ is the orientation angle setting the direction of the active force.
The parameters $\gamma$ and $\gamma_r$ represent the translational and rotational friction coefficients respectively, while $D_t$ and $D_r$ are the corresponding diffusion coefficients. The terms $\boldsymbol{\xi}_i$ and $\eta_i$ denote independent Gaussian white noises with zero mean and unit variance. Each particle, of mass $m$, is self-propelled at a constant speed along the direction $\hat{\mathbf{n}}_i = (\cos\theta_i, \sin\theta_i)$, determined by its orientation angle $\theta_i$. Particle motion is influenced by a chiral term, $\Omega$, which induces circular orbits, and by a deterministic torque, $\mathbf{T}^{\text{sa}}_i = \beta\, (\hat{\mathbf{n}}_i \times \mathbf{v}_i)$, that aligns the orientation $\hat{\mathbf{n}}_i$ with the velocity $\mathbf{v}_i$. The self-alignment strength is set by the parameter $\beta$, which defines the typical distance, $\gamma_r/\beta$, that a particle travels before its orientation aligns with its velocity. 

The force $\mathbf{F}_i$ accounts for purely repulsive interactions between particles and is derived from a Weeks–Chandler–Andersen (WCA) potential: $\mathbf{F}_i = -\nabla_i U_{\rm tot}$, with $U_{\rm tot} = \sum_{i < j} U(|\mathbf{r}_i - \mathbf{r}_j|)$ and $U(r) = 4\epsilon\left[(\sigma/r)^{12} - (\sigma/r)^6\right]$ if $r < 2^{1/6}\sigma$ and zero otherwise. Here, $\epsilon$ sets the energy scale, and $\sigma$ denotes the particle diameter. We consider a regime of high packing fraction $\phi=N \sigma^2\pi/4L^2 \approx 1.1$, where $L$ is the side of the simulation box. In two dimensions, the high value of $\phi$ forces the particles to occupy the vertices of a triangular lattice (Fig.~\ref{fig:figura1}(a)) arranging in an almost perfect crystal-like configuration. As a consequence, they cannot exchange positions with their neighbors. Each individual unit is confined in a cage formed by the nearest particles, and the only way a particle can move beyond the typical interparticle distance is through collective motion. The high density considered in our work allows us to reproduce conditions relevant to many biological systems, such as epithelial tissues~\cite{SawDoostmohammadiEtAl:2017DefectsExtrusion} or chiral cellular aggregates \cite{wang2022cellchirality} while avoiding effects like density fluctuations~\cite{KurodaMatsuyamaKawasakiMiyazaki:2023AnomalousABPFluctuations, NishiguchiSano:2015JanusTurbulence} or density inhomogeneities. 

In our simulations, we set $D_t = 0$, as in active matter systems this term is typically much smaller than the effective diffusion arising from activity.
Chirality introduces a characteristic time for a particle to complete a circular orbit in the absence of other effects, given by $1/\Omega$. Self-alignment defines another characteristic time, $\gamma_r/(\beta v_0)$, i.e., the time required for the particle's orientation to align with its velocity. These times compete with other intrinsic time scales of the system, namely the persistence time of the particle, $\tau = 1/D_r$, and the translational inertial time, $\tau_d = m/\gamma$.

Simulations are performed in a box of side $L$ with periodic boundary conditions by rescaling lengths by the particle diameter $\sigma$ and time by $\tau$. With these choices, the dynamics are governed by several dimensionless parameters: the P\'eclet number, $\text{Pe} = v_0/(D_r \sigma)$, which quantifies the persistence length of a particle relative to its size; the reduced mass, $M = D_r m/\gamma$, which determines the relevance of inertia and is chosen to be small, $M = 10^{-3}$; the reduced chirality, $\omega = \Omega \tau$, which compares the typical time for a particle to complete a circular orbit with the persistence time; the reduced self-alignment strength, $B = \beta \sigma / \gamma_r$, which compares the self-aligning length to the particle size; and, finally, the reduced interaction strength, $\sqrt{\epsilon/m}/(D_r \sigma)$. Details of the numerical simulations are discussed in Appendix~\ref{appendix:A}.


\begin{figure}[t] 
    \centering
    \includegraphics[width=0.92\columnwidth]{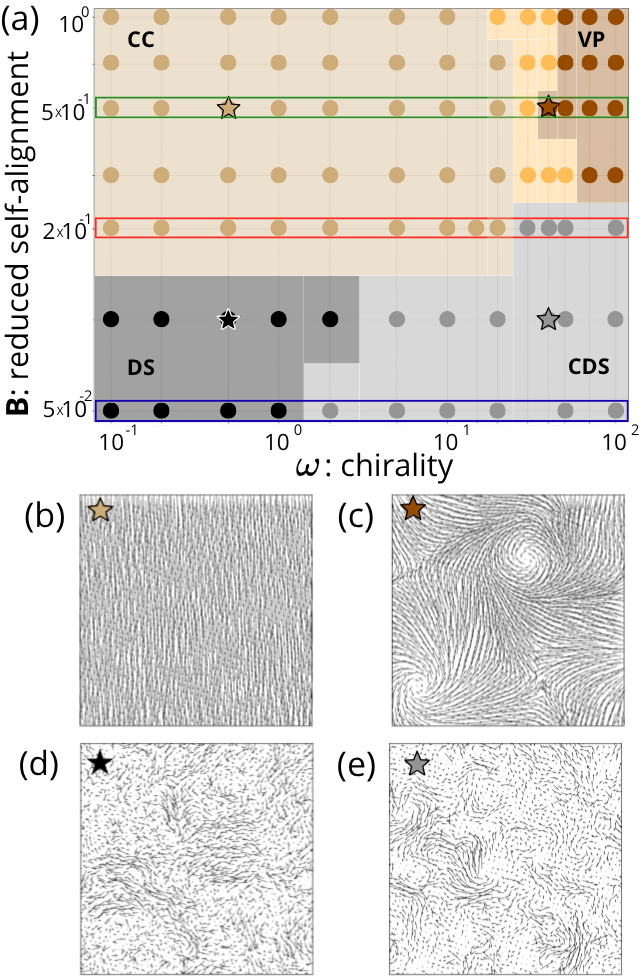}
    \caption{\textbf{Phase diagram.} (a) Phase diagram in the plane of reduced self-alignment strength $B$ and reduced chirality $\omega$. Colors denote different phases which are visualized by the four snapshots (b)–(e), where vectors represent the particle velocities. (b) Circling crystal (CC) phase (light brown), where all particle velocities are aligned, and all the system performs synchronized circular orbits in space. (c) Vortex phase (VP), where the system develops vortex structures and the global circling is suppressed. (d) Disordered system (DS), where particle velocities form finite-size domains where particles are aligned. (e) Chiral disordered system (CDS), where the particles velocity are aligned in finite-size domains, but the particles show oscillatory dynamics. The orange region that separate CC and VP phases identify metastable configurations, showing circling crystal or vortex structures depending on the initial conditions. Colored rectangles in the phase diagram outline the values of parameters analyzed in successive figures.
    The remaining dimensionless parameters of the simulations are $\text{Pe} = 10$, $M = 10^{-4}$, $\sqrt{\epsilon/m}/(D_r \sigma)=10^2$, and $\phi = N \pi \sigma^2 / 4L^2 = 1.1$.}
    \label{fig:figura2}
\end{figure}

\section{Results}
\label{sec:results}

In order to investigate the interplay between chirality and self-alignment in an active crystal, we perform simulations by fixing the P\'eclet number at $\text{Pe} = 10$ and systematically varying the strength of the reduced self-alignment $B$ and the reduced chirality $\omega$ (see Fig.~\ref{fig:figura2}(a)). In Ref.~\cite{Musacchio2025pt}, a non-chiral active solid characterized by self-alignment was already studied and revealed a transition from a disordered to a flocking state as the alignment strength increases. Here, we aim to investigate how the inclusion of a chiral term modifies the dynamics of the crystal, ultimately leading to the complete suppression of the flocking state observed for sufficiently strong chirality.

\subsection{Single-particle dynamics in a  self-aligning chiral active solid}
We start by monitoring the dynamics of a target particle inside the crystal (see the Supplementary Video). The particle trajectories, which in Fig.~\ref{fig:figura1}(b–e) are represented by colored dots that become lighter with time and are connected by gray lines, strongly depend on the relative strengths of reduced chirality $\omega$ and reduced self-alignment $B$. For weak $B$ (i.e., before the flocking transition~\cite{Musacchio2025pt}) and small $\omega$, the particle irregularly moves with a displacement that exceeds the particle diameter $\sigma$ (Fig.~\ref{fig:figura1}(d)). This observation suggests that the system does not flock but the particles move in finite-size domains as in Ref.~\cite{caprini2020hidden}, implying the absence of a global synchronization among the particles of the crystal. In this regime, the increase in chirality induces more localized trajectories (Fig.~\ref{fig:figura1}(e)), with small position fluctuations even smaller than the particle diameter $\sigma$. At low chirality and strong self-alignment (Fig.~\ref{fig:figura1}(b)), particles are able to globally coordinate their motion, giving rise to well-defined circular orbits with a radius that scales as the ratio $v_0 / \Omega$, much larger than the particle diameter $\sigma$. These extended orbits emerge from the collective motion of the system, in which all particles perform a circular coordinate motion, suggesting the occurrence of a new flocking-like state with rotating features. Finally, upon increasing chirality in this regime of strong self-alignment, the radius of the circular orbits decreases (Fig.~\ref{fig:figura1}(c)), becoming even smaller than the particle diameter $\sigma$. The reduced size of the orbits effectively suppresses the fluctuations of the particles' displacement within their lattice cages. However, local alignment seems to be preserved, as reflected by the nearly perfect circular trajectories followed by each particle.

This single-particle trajectory study implies that introducing a sufficiently strong self-alignment enables the particles to coordinate their motion, either globally or locally, depending on the intensity of the chiral term, and effectively reduces noise effects. It is reflected in the almost perfectly circular trajectories shown in Fig.~\ref{fig:figura1}(b,c). 


\subsection{Phase Diagram}

The analysis of the single-particle trajectories suggest the occurrence of collective behavior.
Therefore, we systematically build a phase diagram in the plane of reduced chirality $\omega$ and reduced self-alignment $B$, which allows us to identify four distinct phases (Fig.~\ref{fig:figura2}(a)). For weak values of $B$ and $\omega$, the system remains in a disordered system (DS) phase, where particles are unable to globally coordinate their motion and chirality is too weak to significantly affect the system dynamics: the system shows finite-size domains where particles locally align their velocity (Fig.~\ref{fig:figura2}(d)). This local velocity alignment is a consequence of the high-density and active force persistence and results in short-range velocity correlations, consistently with previous findings without self-alignment and chirality~\cite{caprini2020hidden}. By keeping self-alignment weak and increasing chirality, the system enters the chiral disordered system (CDS) phase. In this regime, particles still exhibit local alignment in the velocity field due to the interplay between active forces and the high packing fraction of the system (Fig.~\ref{fig:figura2}(e)). The main difference between the disordered phase (DS) and the CDS is that increasing the strength of the chiral term modifies the system's dynamics by reducing the size of velocity-aligned domains and introducing an oscillatory behavior in the temporal dynamics. Further details will be discussed in the following sections. 

On the other hand, by increasing the value of the self-alignment strength $B$, the system enters a flocking-like phase typical of crystals composed of self-aligning units (see Ref.~\cite{Musacchio2025pt}), where the particles' velocities are globally aligned. In this regime, while chirality does not alter the transition line from the disordered to the ordered phase, it alters the flock dynamics. Indeed, here, active units do not follow an almost straight trajectory as in the standard case of flocking behavior but display collective circular trajectories, as suggested by tracking a single-particle trajectory inside the crystal (Fig.\ref{fig:figura1}(b)).  
In this regime, all the particles' velocities are aligned (see Fig.~\ref{fig:figura2}(b)) as an effect of self-alignment, but the whole group of particles collectively follow circular orbits due to chirality. Consequently, we term this collective behavior, circling crystal (CC) phase.
The characteristic radius of these circular orbits is determined by the interplay between the active force and chirality and scales as $v_0 / \Omega$. For a fixed self-propulsion of the active units, increasing the value of $\Omega$ leads to a decrease in the typical radius of the orbits, as illustrated by the single-particle trajectories inside the active solid shown in Fig.~\ref{fig:figura2}(b,c).

To theoretically predict the circling crystal phase, we use a method similar to the one used in Ref.~\cite{Musacchio2025pt} by including the additional effect of chirality.
As derived in Appendix~\ref{app:theory}, the particle dynamics (Eqs.~\eqref{eq:dynamics}) can be mapped onto a coarse-grained dynamics in the continuum limit, with the following form
\begin{align}
    \dot{\mathbf{v}}(\mathbf{r},t) &=
    -\frac{1}{\tau}\frac{\delta}{\delta \mathbf{v}(\mathbf{r},t)} 
    \mathcal{F}_{LG}[\mathbf{v}(\mathbf{r},t)] \nonumber \\
    &\quad + \Omega \, \mathbf{z} \times \mathbf{v}(\mathbf{r}, t)
    + v_0 \sqrt{\frac{2}{\tau}} \, \boldsymbol{\eta}(\mathbf{r},t)\,.
\label{eq:dynamics_v}
\end{align}
where $\boldsymbol{\eta}(\mathbf{r},t)$ is a vector representing Gaussian white noise satisfying 
$\langle\boldsymbol{\eta}(\mathbf{r},t)\boldsymbol{\eta}(\mathbf{r}',t')\rangle = \delta(t-t')\delta(\mathbf{r}-\mathbf{r}')$ and $\mathbf{z}$ corresponds to the unit vector normal to the plane of motion.
The term $\mathcal{F}_{LG}[\mathbf{v}(\mathbf{r},t)]$ denotes the Landau–Ginzburg free energy, which reads
\begin{align}
\label{app:eq_landaugizburg_mt}
    \mathcal{F}_{LG}[\mathbf{v}(\mathbf{r},t)] 
&\approx \frac{3\tau\sigma^2}{2}\frac{\mathcal{K}}{\gamma}
\left(\nabla \mathbf{v}(\mathbf{r},t)\right)^2  \nonumber\\
&\hspace{-4em}+ \frac{1}{2} |\mathbf{v}(\mathbf{r},t)|^2
\left(1 - v_0\frac{\tau\beta}{\gamma_r}\right)
+ \frac{\tau\beta}{4 v_0 \gamma_r} |\mathbf{v}(\mathbf{r},t)|^4 .
\end{align}
The first term on the right-hand side of Eq.~\eqref{app:eq_landaugizburg_mt} represents the kinetic contribution, 
which tends to restore random configurations. The remaining two terms correspond to a single-particle 
free energy with a Mexican-hat shape, whose form depends on the sign of the mass term, i.e. the term proportional to $|\mathbf{v}(\mathbf{r},t)|^2$. 
Compared to the results of Ref.~\cite{Musacchio2025pt}, chirality generates an additional force $\Omega \, \mathbf{z} \times \mathbf{v}(\mathbf{r}, t)$ which cannot be expressed as the functional derivative of an effective free energy. The structure of the chirality-induced term is reminiscent of a Lorentz force due to an external magnetic field and therefore it is intuitively responsible for a circular current, applying at the single-particle level. 

To shed light on the effect of the chiral term on the system dynamics, we analyze the complex velocity $\boldsymbol{\psi}(\mathbf{r}, t)=v_x(\mathbf{r}, t)+ i v_y(\mathbf{r}, t)$, which is then convenient to express in polar coordinates  
$\boldsymbol{\psi}(\mathbf{r}, t) = v(\mathbf{r}, t)e^{i\varphi(\mathbf{r}, t)}$, where $v=v(\mathbf{r}, t)$ is the velocity modulus and $\varphi=\varphi(\mathbf{r}, t)$ the polar angle. 
With this choice the dynamics~\eqref{eq:dynamics_v} can be rewritten as
\begin{subequations}
\begin{equation}
\dot{v} = 
- \frac{3 \sigma^2}{2}\frac{\mathcal{K}}{\gamma} v\big(\nabla\varphi\big)^2
- v\left(\frac{1}{\tau}-\beta v_0 \right) 
- \frac{\beta}{v_0 \gamma_r} v^3 \,,
\label{eq:app_dynamicscrystal_time_evol_mod_finale_mt}
\end{equation}
\begin{equation}
\begin{split}
v(\mathbf{r},t)\dot{\varphi} &= 
v(\mathbf{r},t)\Omega 
+ 3 \sigma^2\frac{\mathcal{K}}{\gamma}\nabla v(\mathbf{r}, t)\cdot\nabla\varphi(\mathbf{r},t) \\
&\quad - \frac{3 \sigma^2}{2}\frac{\mathcal{K}}{\gamma} v(\mathbf{r}, t)\nabla^2\varphi(\mathbf{r},t) \,.
\end{split}
\label{eq:app_dynamicscrystal_time_evol_orient_dyn}
\end{equation}
\end{subequations}
The time evolution of the velocity modulus, $v$, is not affected by the chirality and remains consistent with the Mexican-hat free-energy profile described by Eq.~\eqref{app:eq_landaugizburg_mt}. However, we find that the chiral term uniquely influences the time evolution of the velocity orientation $\varphi$ through a constant term. Therefore by considering a reference frame that rotates with angular velocity $\Omega$, i.e.\ implementing the change of variables $\varphi' = \varphi - \Omega t$ in Eq.~\eqref{eq:app_dynamicscrystal_time_evol_orient_dyn}, chirality disappears from the dynamics.
This is equivalent to looking for a solution with a time-shifted phase, which can hold only for weak chirality. In this regime, assuming that spatial gradients of the velocity modulus are negligible, the constantly rotated angle $\varphi'$ is described by a diffusive dynamics
\begin{flalign}
\label{eq:app_dynamicscrystal_time_evol_orient:finale_int_mt}
\dot{\varphi}' =   -  \frac{3 \sigma^2}{2}\frac{\mathcal{K}}{\gamma} \nabla^2\varphi' \,.
\end{flalign}
As a consequence, in the rotating reference frame, the angle field gives rise to the Goldstone modes associated with the spontaneous breaking of rotational symmetry. This theory explains the circling crystal phase because the solution for the polar angle $\varphi$ evolves diffusely but is subject to a constant temporal increase with angular velocity $\Omega$. Since the only effect due to chirality is imposing a constant rotation to the velocity field, our theory predicts that the transition line from the disordered to the circling crystal is not affected by chirality and occurs when the mass term 
changes sign, i.e., when $\beta$ exceeds the critical value $\beta_c$, given by
\begin{equation}
\label{eq:transitionpoint}
    \frac{\beta_c}{\gamma_r} = \frac{1}{v_0 \tau} \, .
\end{equation} 
This theoretical prediction is consistent with our numerical findings (Fig.~\ref{fig:figura2}(a)) and with the flocking transition observed in the absence of chirality~\cite{Musacchio2025pt}.


By maintaining a high value of self-alignment $B$ and selecting strong chirality values $\omega$, the system enters the vortex phase (VP). In this regime, we observe that chirality competes with self-alignment, thereby suppressing the flocking behavior. For $\omega \gtrsim 10$, the orbital radius $r_{\Omega}$ are comparable to the one of the particle, $r_{\Omega} \lesssim \sigma/2$. This indicates that the circular motion of the particles is almost localized within the individual lattice cage (see Fig.~\ref{fig:figura1}(c)), which corresponds to low effective temperature of the system. As a consequence, global alignment -- which generally emerges when self-alignment allows random persistent fluctuations of the particles to break the global rotational symmetry -- is suppressed. Instead, local ordered domains emerge, where within each of the particle velocities are aligned and exhibit coordinated circular motion (see Fig.~\ref{fig:figura2}(c)). To ensure that the VP is not an artificial finite-size effect, we have systematically performed simulations with different particle numbers, verifying that the vortex spacing observed in the velocity field is unaffected by the size of the simulation box.
We remark that, between the circling crystal (CC) phase and the vortex phase (VP), there exists an intermediate region, highlighted in orange in the phase diagram, where the system may show a stable global rotating phase or vortex-like structures, depending on the randomly chosen initial conditions.

\begin{figure}[t] 
    \centering 
    \includegraphics[width=\columnwidth]
    {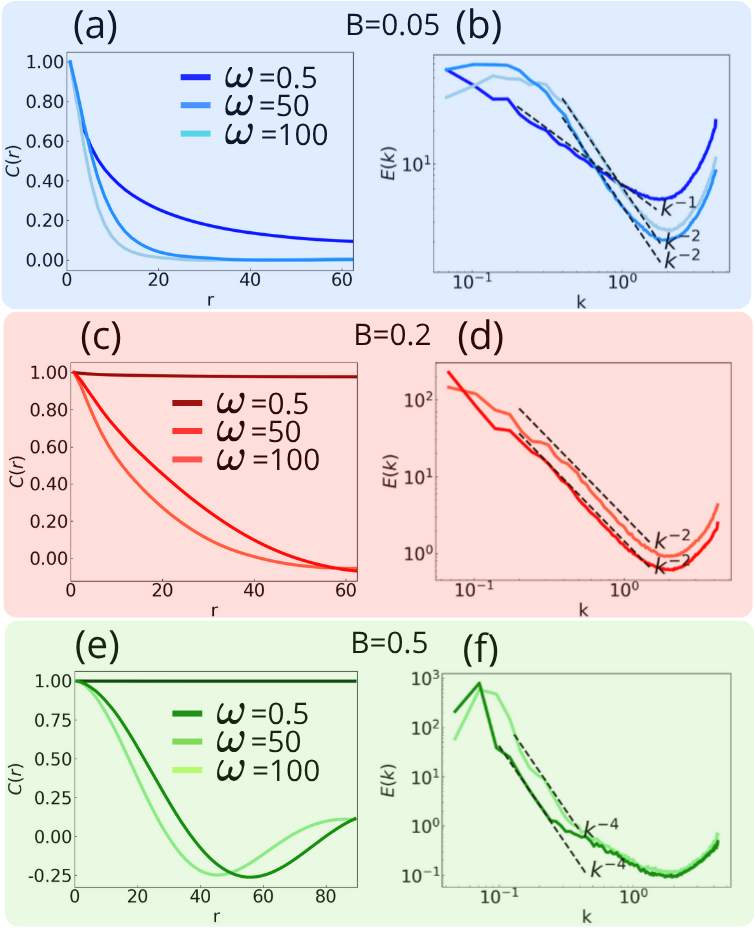}
    \caption{\textbf{Correlation Functions and Energy Spectra.}(a), (c), and (e) Spatial velocity correlations as a function of the distance $r$ for reduced self-alignment values of $B = 0.05$, $0.2$, and $0.5$, respectively. (b), (d), and (f) Kinetic energy spectra of the system, $E(k)$, defined in Eq.~\eqref{eq:energy_spectrum}, as a function of $k$ for $B = 0.05$, $0.2$, and $0.5$. The simulations at $B=0.5$ are performed in a larger box ($180\times180$) in order to avoid finite-size effects. The other dimensionless parameters of the simulations are: $\text{Pe} = 10$, $M = 10^{-4}$, $\sqrt{\epsilon/m}/(D_r \sigma)=10^2$, and $\phi = N \pi \sigma^2 / 4L^2 = 1.1$.}
    \label{fig:figura3}
\end{figure}

\begin{figure*}[t]
    \centering
    \includegraphics[width=0.95\textwidth]{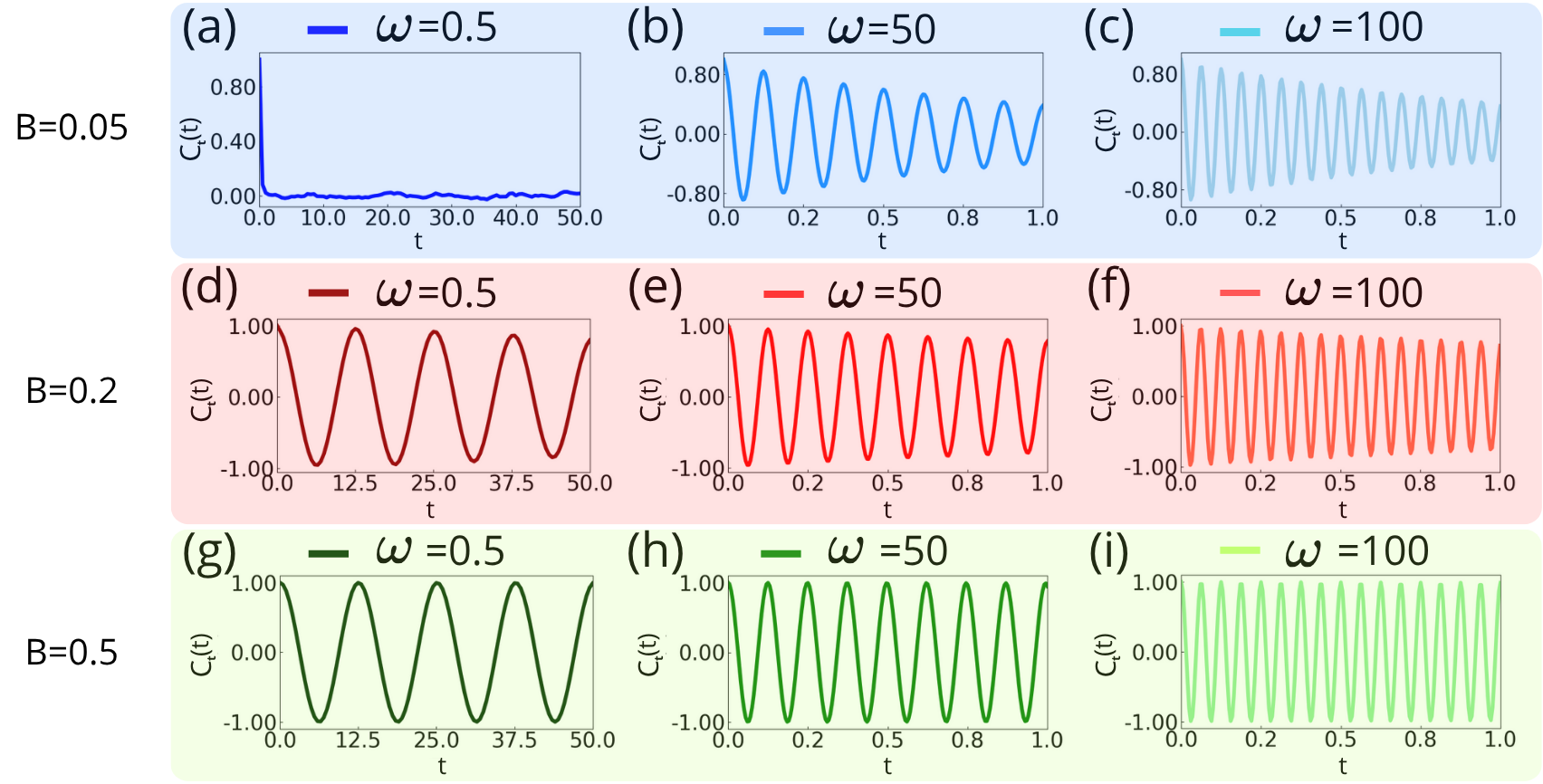}
    \caption{\textbf{Velocity auto-correlation functions.} (a)–(c) Velocity auto-correlation functions $C_t(t)$ for $B = 0.05$ and $\omega$ = 0.5, 50, 100. (d–f) $C_t(t)$ for $B = 0.2$ and $\omega$ = 0.5, 50, 100. (g–i) $C_t(t)$ for $B = 0.5$ and $\omega$ = 0.5, 50, 100. The simulations at $B=0.5$ are performed in a larger box (180×180) in order to avoid finite-size effects. The other dimensionless parameters of the simulations are: $\text{Pe} = 10$, $M = 10^{-4}$, $\sqrt{\epsilon/m}/(D_r \sigma)=10^2$, and $\phi = N \pi \sigma^2 / 4L^2 = 1.1$.}
    \label{fig:figura4}
\end{figure*}

In Fig.~\ref{fig:figura2}(a), three different colors (blue, red, and green) are used to highlight specific lines of the phase diagram corresponding to different values of the reduced self-alignment (\textit{B} = 0.05, 0.2, and 0.5). Along these lines, we observe transitions between all the four phases discussed in this work. The same color scheme will be consistently employed in the subsequent figures.

\subsection{Spatial Velocity Correlation Functions and Energy Spectra}
In order to characterize the phases previously described, we analyzed both the spatial velocity correlation function and the kinetic energy spectrum (Fig.~\ref{fig:figura3}). 
The spatial velocity correlation function is defined as
\begin{equation}
C(r) = \Big\langle \frac{\sum_{ij} \mathbf{v}_i \cdot \mathbf{v}_j \, \delta(r-r_{ij})}{\sum_{ij} \delta(r-r_{ij})} \Big\rangle,
\label{eq:spatial_correlation}
\end{equation}
with $r_{ij} = |\mathbf{r}_i - \mathbf{r}_j|$. Here, the symbol $\langle \cdot \cdot \cdot \rangle$ corresponds to the time average.
In the case of a crystal composed by passive units, $C(r)$ does not have a spatial dependence and is approximately given by a Dirac delta function centered at $r=0$. In the absence of active forces, the particles' velocities are not correlated since the system is governed by the Maxwell-Boltzmann distribution. In an active solid without both chirality and self-alignment, $C(r)$ typically exhibits a spatial decay with an exponential profile and a correlation length increasing with the persistence time $\tau$ of active force~\cite{caprini2020hidden}. When chirality is included in the dynamics of individual particles, the characteristic length of the correlated domains decreases as chirality increases consistently with theoretical and numerical results of Ref.~\cite{Shee2024Emergent, marconi2025spontaneous} in the absence of self-alignment. On the other hand, for strong enough self-alignment the system undergoes a flocking transition. In this case, the velocity correlation function has an almost constant profile~\cite{Musacchio2025pt}.

To focus on the spread of information across scales, we study the kinetic energy spectrum, defined as 
\begin{equation}
E(k) = 2\pi L^{2} k \frac{\langle |\hat{\mathbf{v}}(k)|^{2} \rangle}{\langle v^{2} \rangle}\,.
\label{eq:energy_spectrum}
\end{equation}
Here, $k = |\mathbf{k}|$ corresponds to the modulus of the wave vector and $\hat{\mathbf{v}}(k) = \int d^{2}\mathbf{r} \, \mathbf{v}(\mathbf{r}) \, e^{-i \mathbf{k} \cdot \mathbf{r}}$ coincides with the Fourier transform of the velocity field. The maximum and minimum accessible values of $k$ are defined respectively by the inverse of the particle diameter and by the inverse of the box size, both scaled by $2\pi$. 
In general, when the energy spectra exhibit a power-law decay in the range of wavenumbers $k$ between the value corresponding to the spectral peak and the smallest $k$ defined by the inter-particle distance, this indicates the emergence of a scale-free regime in real space. The presence of a peak at a well-defined wavenumber $k$ signals the existence of structures with a characteristic size in the velocity field. When the peak is not visible, it occurs at smaller values of $k$ that are not accessible due to the finite system size, which imposes a lower limit on the range of $k$ values studied. The increase of the energy spectrum always observed at large $k$ values originates from the intrinsically disordered packing at short length scales.
For a non-chiral active crystal without self-alignment, the energy spectrum profile scales as $E(k)\sim k^{-1}$ for large $k$, consistently with an Ornstein–Zernike profile~\cite{caprini2020spontaneous, henkes2020dense}. In Ref.~\cite{Shee2024Emergent}, where chirality is introduced at the level of single-particle dynamics, 
the energy spectra reveal the emergence of correlated velocity fields for low noise and weak chirality.

In our study, we combine both chirality and self-alignment in a crystal composed by active units. For $B = 0.05$ (Fig.~\ref{fig:figura3}(a,b), blue panel), the spatial velocity correlation functions for $\omega = 0.5, 50,$ and $100$ (Fig.~\ref{fig:figura3}(a)) exhibit an exponential decay, with a characteristic length that decreases as chirality increases. This reflects the fact that, as chirality increases, particles tend to move along orbits of smaller radius, resulting in a system characterized by short range correlations in the velocity field. The corresponding energy spectra (Fig.~\ref{fig:figura3}(b)) show a slower decay at larger $k$ for $\omega = 0.5$ (DS phase), characterized by a power-law behavior of the form $E(k) \sim k^{-1}$, in accordance with an Ornstein–Zernike profile, as discussed in recent works on chiral and non-chiral active crystals \cite{henkes2020dense, Shee2024Emergent, marconi2025spontaneous}. As chirality increases (CDS phase), the spectral peak shifts to larger $k$, reflecting the emergence of smaller structures in the velocity field (Fig.~\ref{fig:figura2}(e)), while the power-law decay becomes steeper and, for extreme chirality strength, scales as $E(k) \sim k^{-2}$. Increasing chirality leads to smaller particle domains where velocities are correlated, but also to a progressive loss of local fluctuations and instabilities, which is reflected in a faster decay of the energy spectrum.

At $B = 0.2$ (Fig.~\ref{fig:figura3}(c,d), red panel) and $\omega = 0.5$, the system enters the circling crystal (CC) phase: all particles are aligned and move collectively along circular orbits (see Fig.~\ref{fig:figura2}(b)), resulting in a nearly constant  spatial velocity correlation function close to one as shown in Ref.~\cite{Musacchio2025pt}. The corresponding energy spectrum displays a delta-like peak at $k = 0$, which cannot be properly extrapolated from the data since the minimum accessible $k$ is determined by the inverse of the simulation box size, scaled by $2\pi$. Increasing $\omega$ to $50$ or $100$ drives the system back into the chiral disordered (CDS) phase, where correlations decay exponentially, and the energy spectrum follows a power-law behavior of the form $E(k) \sim k^{-2}$.

For $B = 0.5$ (Fig.~\ref{fig:figura3}(e,f), green panel), the CC phase persists at $\omega = 0.5$ with similar correlation and spectral features. At higher chirality ($\omega = 50, 100$), the system transitions to the vortex phase (VP), characterized by a correlation function that crosses zero and becomes negative at scales set by the domain size (Fig.~\ref{fig:figura2}(c)), and energy spectra with pronounced peaks at finite $k$. Both the zero-crossing in the spatial velocity correlation function and the peak positions in the spectrum indicate that the characteristic size of coherent velocity structures decreases with increasing chirality. In this case, the energy spectrum at large $k$ scales as a power law, $E(k) \sim k^{-4}$, indicating an almost complete suppression of local instabilities in favor of well-defined vortex structures in the velocity field.

The fast power-law decay observed in the energy spectra for all regimes characterized by high chirality reflects the fact that this term introduces a characteristic correlation length in the velocity field, defining the spatial scale on which most of the energy is concentrated. 

\subsection{Velocity Autocorrelation Function}
To characterize the dynamics of the phases described in the phase diagram, we have also analyzed the velocity autocorrelation function, defined as: 
\begin{equation}
C_t(t) = 
\frac{\langle \mathbf{v}_i(0) \cdot \mathbf{v}_i(t) \rangle}
{\langle \mathbf{v}_i(0) \cdot \mathbf{v}_i(0) \rangle}\,,
\label{eq:autocorrelation}
\end{equation}
where $\langle \cdot \cdot \cdot \rangle$ corresponds to the time and particle average. Starting from the lowest self-alignment value (Fig.~\ref{fig:figura4}(a,b,c), blue panel), $C_t$ for $\omega = 0.5$ (corresponding to the DS phase in Fig.~\ref{fig:figura4}(a)), exhibits a rapid exponential decay. This behavior indicates that the particle dynamics in this regime are dominated by the rotational noise, which tends to randomize the particle orientation over the characteristic time scale $\tau$, in agreement with previous studies on non-chiral active solids. In contrast, for $\omega = 50, 100$ (Fig.~\ref{fig:figura4}(b,c)) we are in CDS phase and the velocity autocorrelation functions display clear oscillatory profiles, with a periodicity set by the chiral term $\Omega$, and an evident damping behavior as also shown in Ref.~\cite{Shee2024Emergent}. These chirality-induced oscillations are consistent with the oscillating mean-square displacement encountered even in a free chiral active particle \cite{Teeffelen2008, Sevilla2016, chan2024chiral} and can be related to the occurrence of the odd diffusive behavior typical of chiral systems~\cite{hargus2021odd, Kalz2022Collisions, Caprini2025ActiveThermodynamics}.

For higher self-alignment values, $B = 0.2$ (Fig.~\ref{fig:figura4}(d,e,f), red panel) and $B = 0.5$ (Fig.~\ref{fig:figura4}(g,h,i), green panel), all velocity autocorrelation profiles, even the ones at small chirality, exhibit oscillatory behavior consistent with the chirality-induced periodicity. 
Furthermore, increasing the self-alignment strength reduces the damping of these oscillations, until, for the strongest self-alignment, the curves no longer display any decay. Indeed, strong self-alignment allows the particles to coordinate their motion in an extremely persistent manner, preventing the loss of coordination that is normally induced by rotational noise.

\section{Conclusions}
\label{sec:conclusions}

In this paper, we have studied the interplay between self-alignment and chirality in active crystals. The first mechanism drives the particle orientation to align with the velocity, while the second one induces the particles to perform circular orbits. Consistently with the previous results, for weak chirality and strong self-alignment, a \textit{circling crystal} phase was found where the individual particles synchronize and all move along the same circular orbit.
When chirality is strong, the circular collective motion is suppressed and the system develops vortex-like structures in the velocity field. In this state, the system displays an oscillatory behavior in the spatial velocity correlations and a time-oscillating velocity autocorrelations. 
Further studies could aim to develop a field theory capable of explaining and predicting the 
suppression of this transition numerically observed, for instance by generalizing the theory developed in Ref.~\cite{kalz2024field} to self-aligning chiral systems.


The results presented in this paper suggest that a natural extension of this work would be to study the effects of self-alignment and chirality in systems of active Brownian particles at lower densities. Recent studies on active turbulence~\cite{keta2024emerging, Klamser2025Directed} indicate that the low-density counterpart of the system presented here may exhibit active turbulent-like behavior for strong chirality. Active turbulence \cite{BratanovJenkoFrey:2015NewClassTurbulence, AlertCasademuntJoanny:2022ActiveTurbulence} differs from the usual inertial turbulence observed in fluids at high Reynolds numbers. In the latter case, the interplay between inertial and viscous forces destabilizes the laminar flow, leading to the formation of vortices. The situation is different in active turbulence: here, no external driving is present, but the internal propulsion of the active units is sufficient to induce complex dynamics, giving rise to self-driven flows that can even become chaotic.

Our findings are amenable to experimental validation in both biological systems and in active granular media~\cite{henkes2011active, kumar2014flocking, Koumakis2016, walsh2017noise, barton2017active, scholz2018rotating, baconnier2022selective, LopezCastano2022ChiralityTransitions, Chor2023, Antonov2025SelfSustained}.
In the biological context, it is well established that many cellular assemblies exhibit self-alignment mechanisms~\cite{henkes2011active}, while others display intrinsic chirality~\cite{wang2022cellchirality}. Exploring systems that combine these two features could provide new insights into their collective behaviors.
Regarding granular systems, several experimental realizations already exhibit self-alignment. Examples include Hexbugs~\cite{baconnier2022selective, chor2023manybody, boriskovsky2025probing}, commercial toys characterized by a certain degree of self-alignment, or 3D-printed granular walkers~\cite{casiulis2025geometric}, where the strength of self-alignment can be tuned by adjusting the mass distribution and the position of the center of friction. Including chirality in these systems can be easily achieved by simply breaking the rotational symmetry of the particle body, as demonstrated in light-driven robots propelled by internal vibrations ~\cite{Siebers2023Exploiting}, intelligent robots equipped with internal motors generating horizontal oscillations~\cite{noirhomme2025brainbots} or air-fluidized disks consisting of a set of blades with a constant tilt~\cite{LopezCastano2022ChiralFlow, LopezCastano2022ChiralityTransitions}. The possibility of combining these two ingredients, self-alignment and chirality, in dense systems could enable the experimental observation of the phases described in this work and provide new insights into the interplay between collective motion, chirality, and structural ordering in active matter.


\appendix

\section{Details on the numerical simulations}
\label{appendix:A}

The dynamics described in the main text by Eqs.~\eqref{eq:eulero_pos} and~\eqref{eq:eulero_ang} are numerically implemented by using the Euler integration scheme with a time step of $\Delta t = 5 \times 10^{-6} \, \tau$. Rescaling time by the persistence time of the particle's trajectory, $\tau$, and positions by the particle's diameter, $\sigma$, the integration scheme is as follows:
\begin{subequations}
\begin{align}
\mathbf{r}'_i(t' + \Delta t') &= \mathbf{r}'_i(t') + \Delta t' \mathbf{v}'_i(t') \label{eq:x} \\[2mm]
\mathbf{v}'_i(t' + \Delta t') &= 
\begin{aligned}[t]
& \mathbf{v}'_i(t') - \frac{\gamma}{m D_r} \Delta t' \gamma \mathbf{v}'_i (t') \\
&\hspace{-4em} + \Delta t' \frac{1}{\sigma m {D_r}^2} \mathbf{F'}_i (\mathbf{r}'_i(t')) 
+ \frac{\gamma}{m D_r} \Delta t' v'_0 \mathbf{\hat{n}}_i
\end{aligned} \label{eq:v} \\[2mm]
\theta'_i (t' + \Delta t') &= 
\begin{aligned}[t]
& \Omega\tau + \frac{\beta \sigma}{\gamma_r} \Delta t' (\mathbf{\hat{n}}_i \times \mathbf{v'_i}) \cdot \hat{\mathbf{e}}_z \\
& + \sqrt{2 \Delta t'} Y_i \,
\end{aligned} \label{eq:omega}
\end{align}
\end{subequations}

In the equations above, the prime symbol denotes dimensionless variables used in our simulations. In this formulation, the particle orientation is given by $\hat{\mathbf{n}}_i = (\cos \theta_i, \sin \theta_i)$, and $Y_i$ are Gaussian random variables with zero mean and unit variance. The force arising from interactions can be written as $\mathbf{F}_i = - \nabla_i U_{\rm tot}$, where $U_{\rm tot} = \sum_{i<j} U(|\mathbf{x}_i - \mathbf{x}_j|)$, with $U$ chosen as a WCA potential, as described in the main text.

The particle dynamics are controlled by the dimensionless parameters discussed in the main text, that we summarize again here: the Péclet number, $\text{Pe} = v_0 / (D_r \sigma)$, that is fixed to $\text{Pe} = 10$ in our numerical study; the reduced mass, $M = D_r m / \gamma$, which compares the inertial time $m/\gamma$ with the persistence time $\tau = 1/D_r$. This parameter corresponds to the inverse of the coefficient multiplying the second and last terms on the right-hand side of Eq.~\eqref{eq:v}. This is fixed to $M = 10^{-4}$ to explore the low-inertia regime; and the dimensionless interaction strength, $\sqrt{\epsilon/m}/(D_r \sigma)$, which can be extracted from the third term of the same equation. 

Chirality and self-alignment introduce additional dimensionless parameters: the reduced chirality, $\omega = \Omega \tau$, varied between $10^{-1} \leq \omega \leq 10^2$, and the reduced self-alignment strength, $B = \beta \sigma / \gamma_r$, varied in the range 5$ \times 10^{-2} \leq B \leq 10^0$. Simulations are performed for a total time of $2 \times 10^{2} \, \tau$ in a square box of size $L = 125$ with periodic boundary conditions. Simulations are performed with a number of particles $N = 2.2 \times 10^4$, yielding a packing fraction of $\Phi = N \pi \sigma^2 / 4 L^2 = 1.1$.

\section{Derivation of the theoretical prediction of the circling crystal}
\label{app:theory}

\subsubsection{Algebraic derivation of the chiral self-aligning active dynamics}

The dynamics for chiral active Brownian particles interacting with self-alignment in two dimensions can be expressed in Cartesian components as follows
\begin{subequations}
\begin{flalign}
\label{eq:over_dotx}
& \dot{\mathbf{v}}_i = \frac{\mathbf{F}_i}{m} - \frac{\gamma}{m}{\mathbf{v}}_i + \frac{\gamma}{m}v_0\hat{\mathbf{n}}_i \\
\label{eq:over_dotn}
& \dot{\hat{\mathbf{n}}}_i =
\begin{aligned}[t]
& -\frac{\hat{\mathbf{n}}_i}{\tau} + \sqrt{\frac{2}{\tau}}\boldsymbol{\eta}_i\times \hat{\mathbf{n}}_i \\
& + \beta (\hat{\mathbf{n}}_i \times {\mathbf{v}}_i) \times \hat{\mathbf{n}}_i + \Omega  \mathbf{z} \times \hat{\mathbf{n}}_i \,.
\end{aligned}
\end{flalign}
\end{subequations}
where $\mathbf{v}_i=\dot{\mathbf{x}}_i$. Here, we have neglected the translational noise to be consistent to the model used for the simulations.  

As a first step, we rewrite the self-alignment interactions in the following form (as already done in Ref.~\cite{Musacchio2025pt})
\begin{equation}
\beta (\hat{\mathbf{n}}_i \times \mathbf{v}_i) \times \hat{\mathbf{n}}_i = \beta \left( \mathbf{v}_i - \hat{\mathbf{n}}_i [\mathbf{v}_i \cdot \hat{\mathbf{n}}_i] \right) \,,
\end{equation}
where we have used that $\mathbf{n}_i^2=1$.
At first, we define the acceleration $\mathbf{s}_i=\dot{\mathbf{v}}_i$ and then apply the time derivative to Eq.~\eqref{eq:over_dotx}. By using Eq.~\eqref{eq:over_dotn} to calculate $\dot{\hat{\mathbf{n}}}_i$, we obtain
\begin{multline} \label{eq:dyn_approx_0}
\dot{\mathbf{s}}_i = -\frac{\gamma}{m}\mathbf{s}_i -\mathbf{v}_j \cdot \nabla_i \nabla_j \frac{U_{tot}}{m} 
+ \frac{\gamma v_0}{m} \Bigg[ -\frac{\hat{\mathbf{n}}_i}{\tau} 
+ \sqrt{\frac{2}{\tau}} \boldsymbol{\eta}_i\times \hat{\mathbf{n}}_i \\
+ \beta \bigg( \mathbf{v}_i - \hat{\mathbf{n}}_i [\mathbf{v}_i \cdot \hat{\mathbf{n}}_i] \bigg) 
+ \Omega \mathbf{z} \times \hat{\mathbf{n}}_i \Bigg] \,.
\end{multline}
In order to express the dynamics as a function of position, velocity, and acceleration only, we eliminate $\mathbf{n}_i$ by using Eq.~\eqref{eq:over_dotx}: 
\begin{align} \label{eq:dyn_approx}
\dot{\mathbf{s}}_i &= -\left(\frac{\gamma}{m} + \frac{1}{\tau} \right)\mathbf{s}_i 
- \mathbf{v}_j \cdot \nabla_i \nabla_j \frac{U_{tot}}{m} 
+ \frac{\mathbf{F}_i}{m \tau} \nonumber \\ 
&\quad + \frac{\gamma v_0}{m}\sqrt{\frac{2}{\tau}} \boldsymbol{\eta}_i\times \hat{\mathbf{n}}_i 
+ \frac{\gamma}{m} \mathbf{v}_i \left(-\frac{1}{\tau} + v_0 \beta \right) \nonumber \\
&\quad - \frac{\gamma  v_0}{m} \beta \Bigg[ \left( \frac{m}{\gamma v_0}\mathbf{s}_i +\frac{\mathbf{v}_i}{v_0} - \frac{\mathbf{F}_i}{\gamma v_0} \right) \nonumber \\
&\quad \Big(\mathbf{v}_i \cdot \left(  \frac{m}{\gamma v_0}\mathbf{s}_i +\frac{\mathbf{v}_i}{v_0} - \frac{\mathbf{F}_i}{\gamma v_0} \right) \Big) \Bigg] \nonumber \\
&\quad + \frac{\gamma v_0}{m}\Omega \mathbf{z} \times\Big(\frac{m}{\gamma v_0}\mathbf{s}_i + \frac{\mathbf{v}_i}{v_0} - \frac{\mathbf{F}_i}{\gamma v_0} \Big) \,.
\end{align}
The dynamics~\eqref{eq:dyn_approx} are equivalent the equations of motion~\eqref{eq:over_dotx} and~\eqref{eq:over_dotn} numerically implemented and result from a change of variables from $(\mathbf{x}_i, \mathbf{v}_i, \hat{\mathbf{n}}_i)$ to $(\mathbf{x}_i, \mathbf{v}_i, \mathbf{s}_i)$.
We remark that chirality solely appears in the last line of Eq.~\eqref{eq:dyn_approx}, while the remaining terms are simply due to the interplay between self-alignment activity and potential forces.

\subsubsection{Vanishing inertia limit}

In the limit of small inertia, the leading contributions to the dynamics~\eqref{eq:dyn_approx} are the terms proportional to $\gamma/m$, and neglecting higher-order contributions we obtain
\begin{align}
\dot{\mathbf{s}}_i &= -\frac{\gamma}{m} \mathbf{s}_i 
- \mathbf{v}_j \cdot \nabla_i \nabla_j \frac{U_{tot}}{m} 
+ \frac{\mathbf{F}_i}{m \tau}  
+ \frac{\gamma v_0}{m}\sqrt{\frac{2}{\tau}} \boldsymbol{\eta}_i\times \hat{\mathbf{n}}_i \nonumber \\
&\quad + \frac{\gamma}{m} \mathbf{v}_i \left(-\frac{1}{\tau} + v_0 \beta \right) \nonumber \\
&\quad - \frac{\gamma  v_0}{m} \beta \Bigg[ \left( \frac{\mathbf{v}_i}{v_0} - \frac{\mathbf{F}_i}{\gamma v_0} \right) 
\Big(\mathbf{v}_i \cdot \left( \frac{\mathbf{v}_i}{v_0} - \frac{\mathbf{F}_i}{\gamma v_0} \right) \Big) \Bigg] \nonumber \\
&\quad + \frac{\gamma v_0}{m}\Omega \mathbf{z} \times 
\Big(\frac{\mathbf{v}_i}{v_0} - \frac{\mathbf{F}_i}{\gamma v_0} \Big) \,.
\label{eq:dyn_approx_over}
\end{align}
In the overdamped limit $m/\gamma \to 0$, one can eliminate the variable $\mathbf{s}_i$, neglecting the time evolution of the acceleration, i.e.\ $\dot{\mathbf{s}_i}=0$. In this way, we obtain an equation of motion for $\mathbf{v}_i$, which reads
\begin{align} 
\mathbf{s}_i = \dot{\mathbf{v}}_i &= -\mathbf{v}_j \cdot \nabla_i \nabla_j \frac{U_{tot}}{\gamma} 
+ \frac{\mathbf{F}_i}{\gamma \tau}  
+ v_0\sqrt{\frac{2}{\tau}} \boldsymbol{\eta}_i\times \hat{\mathbf{n}}_i \nonumber \\
&\quad + \mathbf{v}_i \left(-\frac{1}{\tau} + v_0 \beta \right) \nonumber \\
&\quad - v_0 \beta \Bigg[ \left( \frac{\mathbf{v}_i}{v_0} - \frac{\mathbf{F}_i}{\gamma v_0} \right) 
\Big(\mathbf{v}_i \cdot \left( \frac{\mathbf{v}_i}{v_0} - \frac{\mathbf{F}_i}{\gamma v_0} \right) \Big) \Bigg] \nonumber \\
&\quad + v_0 \Omega \mathbf{z} \times \Big(\frac{\mathbf{v}_i}{v_0} - \frac{\mathbf{F}_i}{\gamma v_0} \Big) \,.
\label{eq:dyn_approx_overover}
\end{align}
The first term in Eq.~\eqref{eq:dyn_approx_overover} is a local alignment term and by calling $N_i$ the instantaneous number of particles interacting with the particle $i$, we have
\begin{align}
\frac{1}{\gamma}\sum_j \nabla_i\nabla_j U_{tot} \cdot \mathbf{v}_j 
&= \frac{1}{\gamma}\Bigg(\sum_{j=1}^{N_i} \nabla_i\nabla_i U(|\mathbf{x}_{ij}|) \cdot \mathbf{v}_i \nonumber \\
&\hspace{-4em} + \sum_{j=1}^{N_i} \nabla_i\nabla_j U(|\mathbf{x}_{ij}|) \cdot \mathbf{v}_j \Bigg) \nonumber \\
&\hspace{-4em} =\frac{1}{\gamma}\sum_{j=1}^{N_i} \nabla_i\nabla_i U(|\mathbf{x}_{ij}|) \cdot \left(\mathbf{v}_i - \mathbf{v}_j \right) \,.
\label{eq:interaction_term}
\end{align}
In this interaction, the particle's velocity $v_i$ is attracted to the velocity of particle $j$ depending on the second derivative of the potential.
The other terms in Eq.~\eqref{eq:dyn_approx_overover} are a force $\mathbf{F}_i$ normalized by the persistence time $\tau$, the chiral term and a noise one. In addition, self-alignment induces other single-particle non-linear terms that depend on $\mathbf{v}_i$ and $\mathbf{F}_i$. 

\subsubsection{Lattice approximation}
In the crystalline configuration, particles occupy the sites of an ordered triangular lattice and exhibit small fluctuations around their equilibrium positions. To proceed with the derivation, we adopt the lattice approximation, effectively freezing the particle positions at their lattice sites.
Under this assumption, the net force on each particle vanishes, $\mathbf{F}_i=0$. In addition, the second derivative of the potential becomes constant and can be expressed as
\begin{equation}
\label{app:condition}
\sum_{j=1}^{n_i} \nabla_i\nabla_i U(|\mathbf{x}_{ij}|) \cdot \left(\mathbf{v}_i -  \mathbf{v}_j \right) = \sum_{j=1}^{N_i} \boldsymbol{\mathcal{H}}_{j}\cdot \left(\mathbf{v}_i -  \mathbf{v}_j \right) \,.
\end{equation}
Here, $N_i = 6$ for a two-dimensional triangular lattice, and the constant matrix $\boldsymbol{\mathcal{H}}_j$, which depends on the distance between particles $i$ and $j$, is obtained by computing the second spatial derivatives of the potential $U$:
\begin{align}
\label{eq:supp_elementH}
\nabla^{\alpha}_i \nabla^{\beta}_i U\left( r_{ij} \right) &= \left[ U''(r_{ij}) + \frac{U'(r_{ij})}{|r_{ij}|} \right] \frac{r_{ij}^{\alpha}r_{ij}^{\beta}}{|r_{ij}|^2} \nonumber \\ 
&\hspace{-1em} - \delta_{\alpha\beta} \frac{U'(r_{ij})}{|r_{ij}|} \,,
\end{align}
where $r_{ij}^{\alpha} = r_i^{\alpha} - r_j^{\alpha}$, with $\alpha=x, y$. In this expression, each prime on the potential $U$ means a spatial derivative.
In addition, the following relation holds:
\begin{equation}
\nabla^{\alpha}_i \nabla^{\beta}_j U= - \nabla^{\alpha}_i \nabla^{\beta}_i U \,.
\end{equation}
The Cartesian components of $r_{ij}^{\alpha}/|\mathbf{r}_{ij}|$ can be written as $\cos{\left(\delta_j\right)}$ and $\sin{\left(\delta_j\right)}$, where $\delta_j$ is the angle between the vector $\mathbf{r}_{ij}$ and the horizontal axis. Owing to the regular hexagonal structure of the cluster, this angle can be expressed as $\delta_j = \delta_0 + j\pi/3$, with $j = 0, 1, \ldots, 5$, where $\delta_0$ denotes the orientation of the hexagon with respect to the reference frame, which we set to zero without loss of generality. In this way, we finally obtain the elements of the matrix $\boldsymbol{\mathcal{H}}$
\begin{align}
\mathcal{H}_{xx}(\sigma)&=U''(\sigma)\cos^2\left(j \frac{\pi}{3}\right)+\frac{U'(\sigma)}{\sigma}\sin^2\left(j \frac{\pi}{3}\right)\\
\mathcal{H}_{yy}(\sigma)&=U''(\sigma)\sin^2\left(j \frac{\pi}{3}\right) +\frac{U'(\sigma)}{\sigma}\cos^2\left(j \frac{\pi}{3}\right) \\
\mathcal{H}_{xy}(\sigma)&= H_{yx}(\sigma)=\left(U''(\sigma) - \frac{U'(\sigma)}{\sigma}\right) \nonumber \\
& \cos\left(j \frac{\pi}{3}\right)\sin\left(j \frac{\pi}{3}\right)\,,
\end{align}
where we have suppressed the explicit dependence on $j$.
Finally, we remark that the sum over the six neighbors of the out-of-diagonal elements of $\boldsymbol{\mathcal{H}}$ vanishes
\begin{equation}
    \sum_j^* \mathcal{H}_{xy}(\sigma) = 0 
\end{equation}
while the same protocol for the diagonal elements gives a constant $\mathcal{K}$ with the following expression
\begin{align}   
\label{eq:constantK}
 \mathcal{K}=\sum_j^* \mathcal{H}_{xx}(\sigma) = \sum_j^* \mathcal{H}_{yy}(\sigma)= \nonumber \\
 = 3 \left(U''(\sigma) +\frac{U'(\sigma)}{\sigma} \right) \,.
\end{align}
In this way, we have
\begin{equation}
\sum_J^* \boldsymbol{\mathcal{H}}_j = \mathcal{K} \boldsymbol{\mathcal{I}} \,,
\end{equation}
where $\boldsymbol{\mathcal{I}}$ is the identity matrix.
At the end, the particle dynamics in the lattice approximation can be expressed as
\begin{align} \label{eq:app_dynamicscrystal}
\dot{\mathbf{v}}_i = \mathbf{s}_i &= 
- \frac{1}{\gamma} \sum_{j=1}^{n_i} \mathcal{K} \left(\mathbf{v}_i -  \mathbf{v}_j \right) 
- \mathbf{v}_i\left(\frac{1}{\tau}-\beta v_0 \right) \nonumber \\
&\hspace{-2em} - \frac{\beta}{v_0 \gamma_r} \mathbf{v}_i |\mathbf{v}_i|^2 + \Omega \mathbf{z} \times \mathbf{v}_i 
+ v_0\sqrt{\frac{2}{\tau}} \boldsymbol{\eta}_i\times \hat{\mathbf{n}}_i \,.
\end{align}
where the constant $\mathcal{K}$ is defined by Eq.~\eqref{eq:constantK} and we have neglected directional contribution of the lattice. In this way, the first term on the right-hand side of Eq.~\eqref{eq:app_dynamicscrystal} corresponds to a discrete Laplacian.

\subsubsection{Continuum limit}
Here, we consider the continuum limit by replacing $\mathbf{v}_i \to \mathbf{v}(\mathbf{r}, t)$ where $\mathbf{r}$ is a continuous coordinate on the lattice.
In this way, the discrete Laplacian in Eq.~\eqref{eq:app_dynamicscrystal} is replaced by
\begin{equation}
    - \frac{1}{\gamma} \sum_{j=1}^{n_i} \mathcal{K}\left(\mathbf{v}_i -  \mathbf{v}_j \right) \to \frac{3 \sigma^2}{2}\frac{\mathcal{K}}{\gamma}\nabla^2\mathbf{v}(\mathbf{r}, t)\,.
\end{equation}
where the particle diameter $\sigma$ also corresponds to the lattice constant.
The dynamics~\eqref{eq:dyn_approx} in 
the continuum limit becomes a dynamics for the velocity field $\mathbf{v}(\mathbf{r})$:
\begin{align} \label{eq:app_dynamicscrystal_continuum}
\dot{\mathbf{v}}(\mathbf{r},t) &= 
\frac{3 \sigma^2}{2}\frac{\mathcal{K}}{\gamma}\nabla^2\mathbf{v}(\mathbf{r}, t)  
- \mathbf{v}(\mathbf{r}, t)\left(\frac{1}{\tau}-\beta v_0 \right) \nonumber \\
&\quad - \frac{\beta}{v_0 \gamma_r} \mathbf{v}(\mathbf{r}, t) |\mathbf{v}(\mathbf{r}, t)|^2 \nonumber \\
&\quad + \Omega \mathbf{z} \times \mathbf{v}(\mathbf{r}, t)
+ v_0\sqrt{\frac{2}{\tau}} \boldsymbol{\eta} (\mathbf{r}, t) \,.
\end{align}
The same dynamics can be written also in terms of the functional derivative of a free-energy functional
\begin{equation}
    \dot{\mathbf{v}}(\mathbf{r},t)  =-\frac{1}{\tau}\frac{\delta}{\delta \mathbf{v}(\mathbf{r},t)} \mathcal{F}_{LG}[ \mathbf{v}(\mathbf{r},t)]  + \Omega \mathbf{z} \times \mathbf{v}(\mathbf{r}, t) + v_0\sqrt{\frac{2}{\tau}}\boldsymbol{\eta}(\mathbf{r},t)\,.
\end{equation}
where the term $\boldsymbol{\eta}(\mathbf{r},t)$ is a vector representing white noise such that $\langle\boldsymbol{\eta}(\mathbf{r},t)\boldsymbol{\eta}(\mathbf{r}',t')\rangle=\delta(t-t')\delta(\mathbf{r}-\mathbf{r}')$,  $\Omega \mathbf{z} \times \mathbf{v}(\mathbf{r}, t)$ is the term introduces the chirality in the dynamics and $\mathcal{F}_{LG}[\mathbf{v}(\mathbf{r},t)]$ is the Landau-Ginzburg free energy, which reads:
\begin{align} \label{app:eq_landaugizburg}
\mathcal{F}_{LG}[\mathbf{v}(\mathbf{r},t)] 
&\approx \frac{3\tau\sigma^2}{2}\frac{\mathcal{K}}{\gamma}\left(\nabla \mathbf{v}(\mathbf{r},t)\right)^2 \nonumber \\
&\hspace{-2em}+ \frac{1}{2} \sum_i|\mathbf{v}(\mathbf{r},t)|^2\left(1 - v_0\frac{\tau\beta}{\gamma_r}\right) \nonumber \\
&\hspace{-2em} + \frac{\tau\beta}{4 v_0 \gamma_r} |\mathbf{v}(\mathbf{r},t)|^4 \,.
\end{align}
The first term on the right-hand side of Eq.~\eqref{app:eq_landaugizburg} represents the kinetic contribution, which tends to restore random configurations. The two other terms correspond instead to a single-particle free-energy with a Mexican-hat shape, whose form depends on the sign of the mass term.
\subsubsection{Circling Crystal prediction}
At this point, in order to investigate the effect of the chirality, we can express the velocity field as a complex number, $\boldsymbol{\psi} (\mathbf{r},t) = v_x(\mathbf{r},t) + i v_y(\mathbf{r},t)$, and decompose the dynamics into its cartesian components as follows:
\begin{align} \label{eq:app_dynamicscrystal_continuum_vx}
\dot{v}_x(\mathbf{r},t) &= 
\frac{3 \sigma^2}{2}\frac{\mathcal{K}}{\gamma}\nabla^2 v_x(\mathbf{r}, t)  
- v_x(\mathbf{r}, t)\left(\frac{1}{\tau}-\beta v_0 \right) \nonumber \\ 
&\quad - \frac{\beta}{v_0 \gamma_r} v_x(\mathbf{r}, t) |\mathbf{v}(\mathbf{r}, t)|^2 \nonumber \\
&\quad - \Omega v_y(\mathbf{r}, t) 
+ v_0\sqrt{\frac{2}{\tau}} \eta_x (\mathbf{r}, t) \,.
\end{align}

\begin{align} \label{eq:app_dynamicscrystal_continuum_vy}
\dot{v}_y(\mathbf{r},t) &= 
\frac{3 \sigma^2}{2}\frac{\mathcal{K}}{\gamma}\nabla^2 v_y(\mathbf{r}, t)  
- v_y(\mathbf{r}, t)\left(\frac{1}{\tau}-\beta v_0 \right) \nonumber \\
&\quad - \frac{\beta}{v_0 \gamma_r} v_y(\mathbf{r}, t) |\mathbf{v}(\mathbf{r}, t)|^2 \nonumber \\
&\quad + \Omega v_x(\mathbf{r}, t) 
+ v_0\sqrt{\frac{2}{\tau}} \eta_y (\mathbf{r}, t) \,.
\end{align}
Recombining the two equations, we obtain
\begin{align}\label{eq:app_dynamicscrystal_continuum_recombined}
\dot{v}_x(\mathbf{r},t) + i \dot{v}_y(\mathbf{r},t) &= 
\frac{3 \sigma^2}{2}\frac{\mathcal{K}}{\gamma}\nabla^2\big(v_x(\mathbf{r}, t)+iv_y(\mathbf{r}, t)\big) \nonumber \\
&\hspace{-4em} - \big(v_x(\mathbf{r}, t) + iv_y(\mathbf{r}, t)\big)\left(\frac{1}{\tau} -\beta v_0 \right) \nonumber \\
&\hspace{-4em} - \frac{\beta}{v_0 \gamma_r} \big(v_x(\mathbf{r}, t) + iv_y(\mathbf{r}, t)\big) |\mathbf{v}(\mathbf{r}, t)|^2 \nonumber \\
&\hspace{-6em} + i \Omega \big(v_x(\mathbf{r}, t) + iv_y(\mathbf{r}, t)\big) + v_0\sqrt{\frac{2}{\tau}} \boldsymbol{\eta} (\mathbf{r}, t) \,.
\end{align}
Recalling the complex expression for $\boldsymbol{\psi} (\mathbf{r},t) = v_x (\mathbf{r},t) + i v_y (\mathbf{r},t)$ and neglecting the noise term, which is not essential for the following results, we can write the following equation:
\begin{align} \label{eq:app_dynamicscrystal_complex}
\dot{\mathbf{\psi}}(\mathbf{r},t) &= 
\frac{3 \sigma^2}{2}\frac{\mathcal{K}}{\gamma}\nabla^2\mathbf{\psi}(\mathbf{r}, t) - \mathbf{\psi}(\mathbf{r}, t)\left(\frac{1}{\tau}-\beta v_0 \right) \nonumber \\
&\quad - \frac{\beta}{v_0 \gamma_r} \mathbf{\psi}(\mathbf{r}, t) |\mathbf{\psi}(\mathbf{r}, t)|^2 + i\Omega \mathbf{\psi}(\mathbf{r}, t) \,.
\end{align}
where we have used that $|\mathbf{v}(\mathbf{r}, t)|^2=|\mathbf{\psi}(\mathbf{r}, t)|^2$. 
Expressing the complex velocity in its exponential form, $\boldsymbol{\psi}(\mathbf{r}, t) = v(\mathbf{r}, t) e^{i\varphi(\mathbf{r}, t)}$, in terms of the velocity modulus $v(\mathbf{r}, t)$ and the polar angle $\varphi(\mathbf{r}, t)$, and substituting this into Eq.~\eqref{eq:app_dynamicscrystal_complex}, we can obtain evolution equations for $v(\mathbf{r}, t)$ and $\varphi(\mathbf{r}, t)$ by calculating the real and imaginary parts in Eq.~\eqref{eq:app_dynamicscrystal_complex}, obtaining
\begin{align} \label{eq:app_dynamicscrystal_time_evol_mod}
\text{Re: } \dot{v}(\mathbf{r},t) &= 
\frac{3 \sigma^2}{2}\frac{\mathcal{K}}{\gamma}\nabla^2v(\mathbf{r},t) 
- \frac{3 \sigma^2}{2}\frac{\mathcal{K}}{\gamma} v(\mathbf{r}, t)\big(\nabla\varphi(\mathbf{r},t)\big)^2 \nonumber \\
&\hspace{-2em} - v(\mathbf{r}, t)\left(\frac{1}{\tau}-\beta v_0 \right) 
- \frac{\beta}{v_0 \gamma_r} v(\mathbf{r}, t)^3 \,.
\end{align}
\begin{align} \label{eq:app_dynamicscrystal_time_evol_orient}
\text{Im: } v(\mathbf{r},t)\dot{\varphi} &= 
v(\mathbf{r},t)\Omega 
+ 3 \sigma^2\frac{\mathcal{K}}{\gamma}\nabla v(\mathbf{r}, t)\cdot\nabla\varphi(\mathbf{r},t) \nonumber \\
&\hspace{-2em} - \frac{3 \sigma^2}{2}\frac{\mathcal{K}}{\gamma} v(\mathbf{r}, t)\nabla^2\varphi(\mathbf{r},t) \,.
\end{align}
where we have used:
\begin{flalign}
\label{eq:app_dynamicscrystal_time_evol}
\dot{\mathbf{\psi}}(\mathbf{r},t) = & \dot{v}(\mathbf{r},t) e^{i\varphi(\mathbf{r},t)} + i \dot{\varphi} (\mathbf{r},t)v(\mathbf{r},t)\,.
\end{flalign}
Finally, since $v > 0$ and that it does not vary in space, we obtain:
\begin{align} \label{eq:app_dynamicscrystal_time_evol_mod_finale}
\text{Re: } \dot{v}(\mathbf{r},t) &= 
- \frac{3 \sigma^2}{2}\frac{\mathcal{K}}{\gamma} v(\mathbf{r}, t)\big(\nabla\varphi(\mathbf{r},t)\big)^2 \nonumber \\
&\hspace{-1em} - v(\mathbf{r}, t)\left(\frac{1}{\tau}-\beta v_0 \right) 
- \frac{\beta}{v_0 \gamma_r} v(\mathbf{r}, t)^3 \,.
\end{align}

\begin{flalign}
\label{eq:app_dynamicscrystal_time_evol_orient:finale}
\text{Im: } \dot{\varphi} =  \Omega - \frac{3 \sigma^2}{2}\frac{\mathcal{K}}{\gamma}\nabla^2\varphi(\mathbf{r},t). \
\end{flalign}
Integrating this last expression over time, we obtain
\begin{flalign}
\label{eq:app_dynamicscrystal_time_evol_orient:finale_int}
(\varphi - \varphi_0) =  \Omega t -  \frac{3 \sigma^2}{2}\frac{\mathcal{K}}{\gamma} \int dt \, \nabla^2\varphi(\mathbf{r},t).
\end{flalign}
From this result, it is clear that an additional contribution appears in the time evolution of the velocity orientation: the chiral term $\Omega t$, which induces a constant rotation of the particle velocity. At this point, it is convenient to define a new angle $\varphi' = \varphi - \Omega t$, corresponding to a reference frame rotating with angular velocity $\Omega$. In this new referent frame the dynamics of the system remain the same described in Ref. \cite{Musacchio2025pt} and the transition from a disordered to an ordered state occurs when the mass term in Eq.~\eqref{app:eq_landaugizburg} changes sign, i.e., when $\beta$ exceeds the critical value $\beta_c$, given by
\begin{equation}
\label{eq:transitionpoint}
    \frac{\beta_c}{\gamma_r} = \frac{1}{v_0 \tau} \, .
\end{equation}

\section{Supplementary video details}
\label{appendix:C}

The Supplemental Video provided in the Supplementary Material illustrates the temporal evolution of a chiral active crystal subject to self-alignment, as described by Eq.~\eqref{eq:dynamics}. In the video, the velocity of each particle is represented by a black arrow, while the trajectory and behavior of a selected target particle are highlighted in red. Four different time evolutions of the system are shown in parallel, corresponding to distinct values of the self-alignment strength and chirality. The stars displayed in each panel indicate the phases under consideration, and their colors follow the same convention as in Fig.~\ref{fig:figura2}. Specifically, the top and bottom panels correspond to reduced self-alignment parameters of $B = 0.5$ and $B = 0.1$, respectively, while the left and right panels correspond to chirality values of $\omega = 0.5$ and $\omega = 40$.

For $B = 0.1$ and $\omega = 0.5$ (black star) or $\omega = 40$ (grey star), the system exhibits the Disordered System (DS) or Chiral Disordered System (CDS) phases, respectively, where particle velocities display only local alignment. In the CDS phase, the target particle undergoes a periodic rotation induced by the chiral term, as revealed by the trajectory of the tagged particle.
For $B = 0.5$ and $\omega = 0.5$ (beige star), the system enters the Circling Crystal (CC) phase, characterized by global alignment of particle velocities and coherent circular motion of the entire crystal, as evidenced by the trajectory of the target particle.
Finally, for $B = 0.5$ and $\omega = 40$ (brown star), the system reaches the Vortex Phase (VP), where particle velocities remain aligned and coordinated within finite-size domains, and the velocity field exhibits vortex-like structures. The tagged particle still performs a periodic rotation, but due to strong chirality its motion becomes almost localized within the crystal cage.


\section*{Acknowledgments}
HL acknowledges support by the Deutsche Forschungsgemeinschaft (DFG) through the SPP 2265, under grant numbers LO 418/25. 
LC acknowledges financial support from the University of Rome La Sapienza, under the project Ateneo 2024 (RM124190C54BE48D). 

    \bibliographystyle{apsrev4-1}

    \bibliography{bib.bib}





\end{document}